\documentclass[prl,twocolumn,preprintnumbers,amsmath,amssymb]{revtex4}
\usepackage{graphicx}

\begin{document}

\title{  Light scattering detection of quantum phases of ultracold atoms in optical lattices }
\author{     Jinwu Ye $^{1,2}$, J.M. Zhang $^{3}$, W.M. Liu $^{3}$, Keye Zhang $^{4}$, Yan Li$^{4}$  and Weiping Zhang $^{4}$ }
\affiliation{ $^{1}$ Department of Physics, Capital Normal University, Beijing, 100048 China  \\
 $^{2}$ Department of Physics, The Pennsylvania State
University, University Park, PA, 16802, USA   \\
$^{3}$Institute of Physics, Chinese Academy of Sciences,
Beijing, 100080, China   \\
$^{4}$ Department of Physics, East China Normal university,
Shanghai, 200062, China}
\date{\today }

\begin{abstract}
  Ultracold atoms loaded on optical lattices can provide unprecedented
  experimental systems for the quantum simulations and manipulations
  of many quantum phases.  However, so far, how to detect these quantum phases effectively  remains an outstanding challenge. Here,
  we show that the optical Bragg scattering of cold atoms loaded on optical lattices can be used to detect many quantum phases
  which include not only the conventional superfluid and Mott insulating phases, but also other important phases such as
  various kinds of density waves (CDW),  valence bond solids (VBS), CDW supersolids and VBS supersolids.
\end{abstract}
\maketitle

  Various kinds of strongly correlated quantum  phases of matter may
  have wide applications in quantum information processing, storage
  and communications \cite{manybody}.  It was widely believed  and also partially established that
  due to the tremendous tunability of all the parameters in this
  system, ultracold atoms loaded on optical lattices (OL) can provide an unprecedented
  experimental systems for the quantum simulations and manipulations
  of these quantum phases and quantum phase transitions between these phases.
  For example, Mott and superfluid phases \cite{boson} may have been successfully
  simulated and manipulated by ultra-cold atoms loaded in a cubic optical
  lattice \cite{bloch}. However, there are still at least two outstanding problems
  remaining. The first is  how to realize many important
  quantum phases \cite{manybody}. The second is that assuming the favorable conditions to realize these quantum phases
  are indeed achieved in experiments, how to detect them without ambiguity.
  In this paper, we will focus on the second question.
   So far the experimental way to detect these quantum phases is mainly through
   the time of flight (TOF) measurement \cite{manybody,bloch}
   which simply opens the trap and turn off the optical lattice and let the trapped atoms  expand
   and interfere, then take the image.  The atom Bragg spectroscopy is based on  stimulated matter waves scattering by
  {\sl  two } incident laser pulses \cite{braggbog,braggeng} through the TOF measurements. The momentum \cite{braggbog} transfer Bragg spectroscopy
  was used to detect the Bogoliubov mode inside an BEC condensate.
  The energy transfer \cite{braggeng} Bragg spectroscopy
  was used to detect the Mott gap in a Mott state in an optical lattice.
  Optical Bragg scattering (Fig.1) has been used previously
  to study periodic lattice structures of cold atoms loaded on
  optical lattices \cite{braggsingle}.
  It was also proposed as an effective method for the thermometry
  of fermions in an optical lattice \cite{braggthermal} and
  to detect putative anti-ferromagnetic (AF) ground state of fermions in OL \cite{braggafm}.
  There are very recent optical Bragg scattering experimental
  data from a Mott state, a BEC and AF  state \cite{blochlight}.
  The atom Bragg spectroscopy and Optical Bragg scattering are two different, but complementary
  experimental methods.

  In this paper, we will develop a systematic theory of using the optical Bragg scattering ( Fig.1) to detect the nature of
  quantum phases  of  interacting bosons loaded in optical lattices.
   We show that the optical Bragg scattering not only couples to
   the density order parameter, but also the  {\sl valence bond order } parameter due to the hopping of the bosons
   on the lattice. At integer fillings, when $ \vec{q} $ matches a reciprocal lattice vector $ \vec{K} $ of the underlying OL,
   there is an increase in the optical scattering cross section as the system evolves from the Mott to the SF state due to the increase
   of hopping in the SF state.
   At $1/2 $ filling, in the CDW state, when $  \vec{q} $ matches the CDW ordering
   wavevector $ \vec{Q}_n $ and $ \vec{K} $, there is a diffraction peak proportional to the CDW order parameter squared and the density squared
   respectively (Fig.3a), the ratio of the two peaks is a good measure of the CDW order parameter. In the VBS state, when
   $ \vec{q} $ matches the VBS ordering wavevector $ \vec{Q}_K $, there is a much smaller, but detectable diffraction peak
   proportional to the VBS order parameter squared,  when it matches $ \vec{K}
   $, there is also a diffraction peak proportional to the uniform
   density in the VBS state (Fig.3b). All the diffraction peaks scale as the
   square of the numbers of atoms inside the trap.  All these characteristics can determine uniquely CDW and VBS state at $1/2 $ filling and
   the corresponding CDW supersolid and VBS supersolid slightly away from the $1/2$ filling.
   In the following, we just take 2d optical lattices as examples. The 1d and 3d cases can be similarly discussed.

\begin{figure}
\includegraphics[width=3cm]{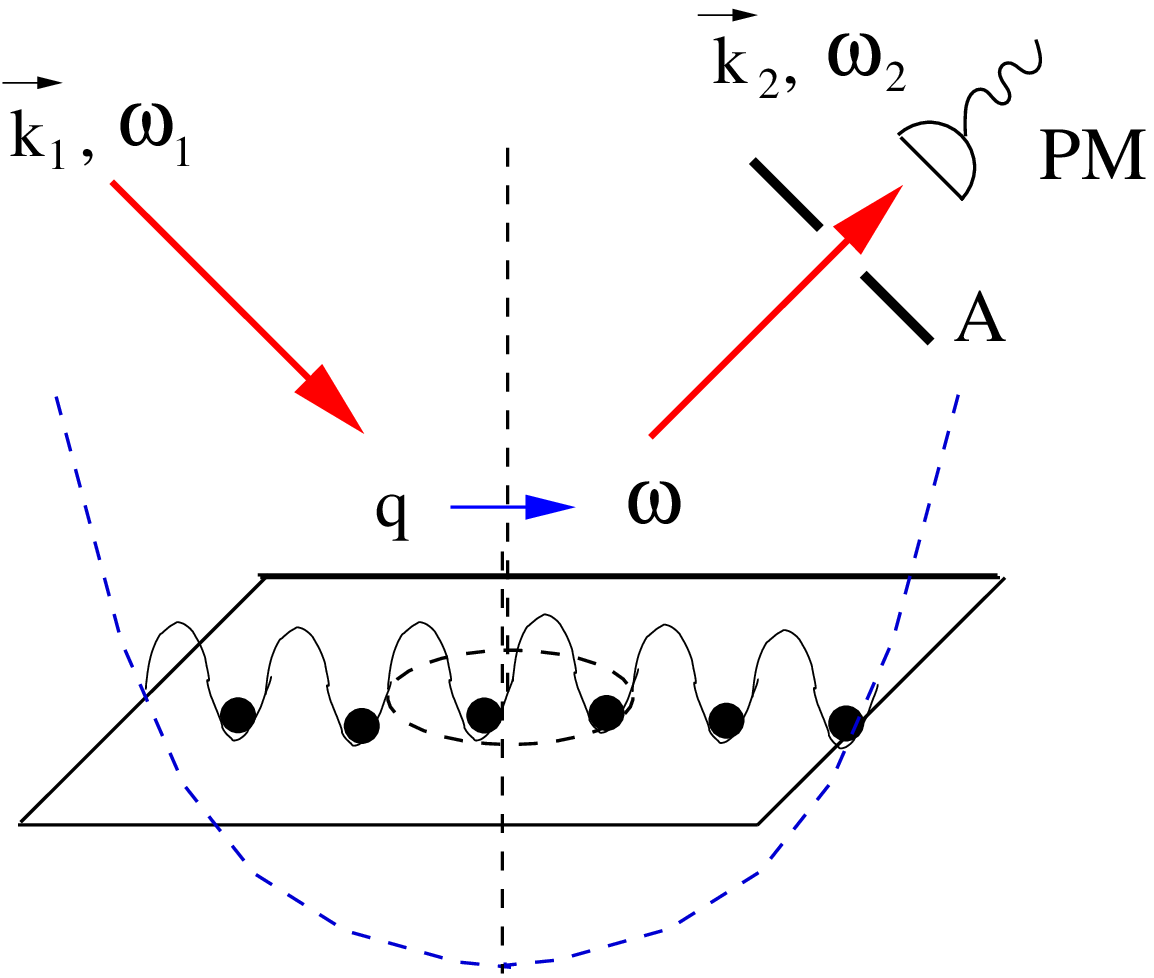}
\hspace{0.5cm}
\includegraphics[width=3cm]{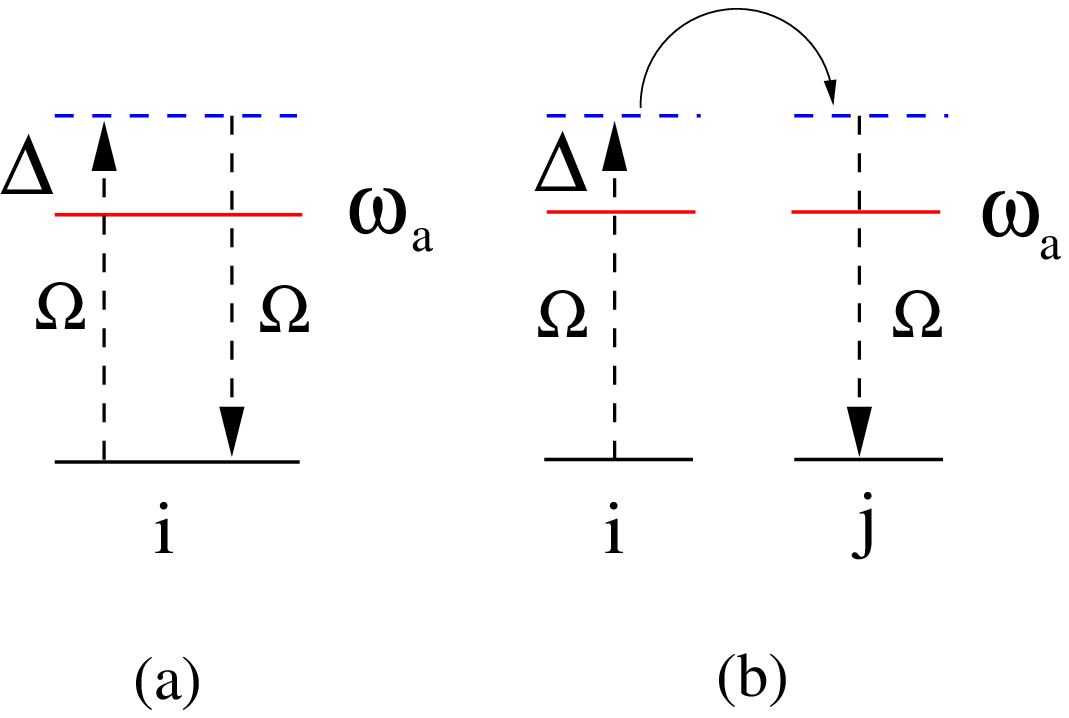}
\caption{ Optical Bragg scattering of cold atoms moving in 2
dimensional optical lattices. The $ \vec{q} =
\vec{k}_{1}-\vec{k}_{2}  $ and
 $ \omega = \omega_1-\omega_2 $ are momentum and energy transfer from the laser beams to the cold atoms respectively.
 The A stands for a aperture, the PM stands for a Photomultiplier.
 The off resonant scattering processes lead to the on-site term (a) and the off-site term (b) in Eqn.3.} \label{fig1}
\end{figure}



    The Extended Boson Hubbard Model (EBHM) with various  kinds of interactions, on all kinds of lattices and at different filling factors
   is described by the following Hamiltonian \cite{boson,gan,square,frusqmc,pq1,mob,bipart,frus,honey}:
\begin{eqnarray}
  H_{BH} & = & -t \sum_{ \langle ij \rangle } ( b^{\dagger}_{i} b_{j} + h.c. ) -
        \mu \sum_{i} n_{i} + \frac{U}{2} \sum_{i} n_{i} ( n_{i} -1 )
                             \nonumber \\
        & +  & V_{1} \sum_{ <ij> } n_{i} n_{j}
        + V_{2} \sum_{ \langle ik \rangle } n_{i} n_{k} + \cdots
\label{boson}
\end{eqnarray}
    where $ n_{i} = b^{\dagger}_{i} b_{i} $ is the boson density, $ t $ is the nearest neighbor hopping which
    can be tuned by the depth of the optical lattice potential, the
    $ U, V_{1}, V_{2} $ are onsite, nearest neighbor (nn) and next nearest neighbor (nnn) interactions respectively,
    the $ \cdots $ may include further neighbor interactions
    and possible ring-exchange interactions. The filling factor $ n= N_a/N $ where $ N_a $ is the number of atoms and $ N $ is the
    number of lattice sites. The on-site interaction
    $ U $ can be tuned by the Feshbach resonance \cite{boson}.
    Various kinds of optical lattices such as honeycomb, tri-
    angular \cite{honeyexp}, body-centered-cubic  \cite{honeyexp}, Kagome lattices \cite{kalattice} can be
    realized by suitably choosing the geometry of the laser beams
     forming the optical lattices.
    There are many possible ways to generate longer range interaction $ V_{1}, V_{2},....$ of
     ultra-cold atoms loaded in optical lattices.
     Being magnetically  or electrically polarized, the $ ^{52}Cr $  atoms \cite{cromium} or
     polar molecules \cite{junpolar} $ ^{40}K+ ^{87} Rb $  ( or $ ^{39} K+ ^{87} Rb $ )
     interact with each other via long-rang anisotropic
     dipole-dipole interactions. Loading the $ ^{52}Cr $ or the polar molecules on a 2d optical lattice
     with the dipole moments perpendicular to the trapping plane can be mapped to
     Eqn.\ref{boson} with long-range  repulsive interactions $ \sim p^{2}/r^{3}
     $ where $ p $ is the dipole moment.  The CDW supersolid phases studied by QMC \cite{square}
     and described in \cite{bipart} by the dual vortex method
     was numerically found to be stable in large parameter regimes  in this system \cite{polarsszoller}.
     The generation of the ring exchange interaction has been discussed in \cite{qs}.
     Some of the important phases with long range interactions
     are listed in Fig.2. Recently, the quantum entanglement properties of the
     VB state was addressed in \cite{vbsent}.

\begin{figure}
\includegraphics[width=3.5cm]{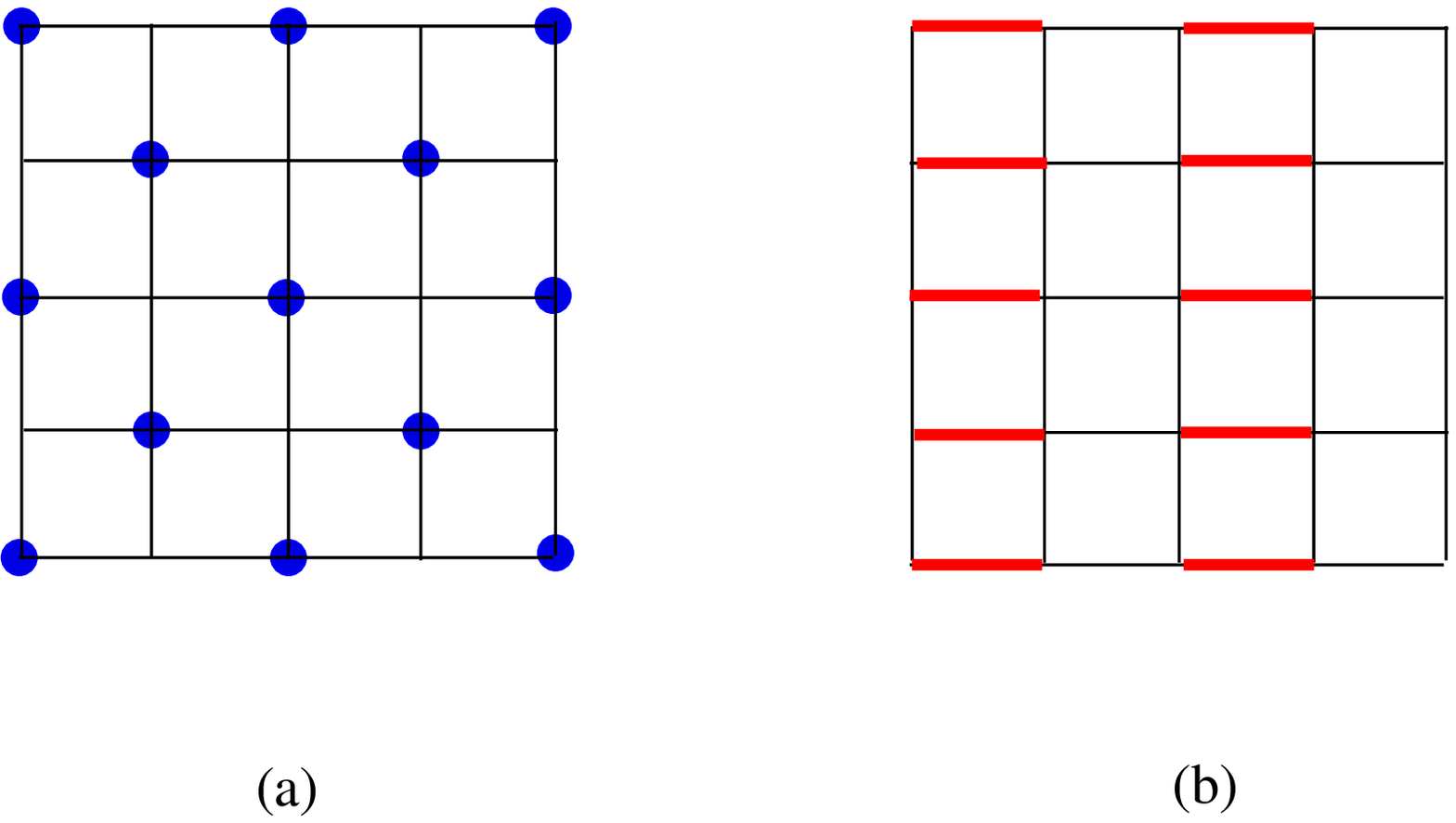}
\hspace{0.5cm}
\includegraphics[width=3.5cm]{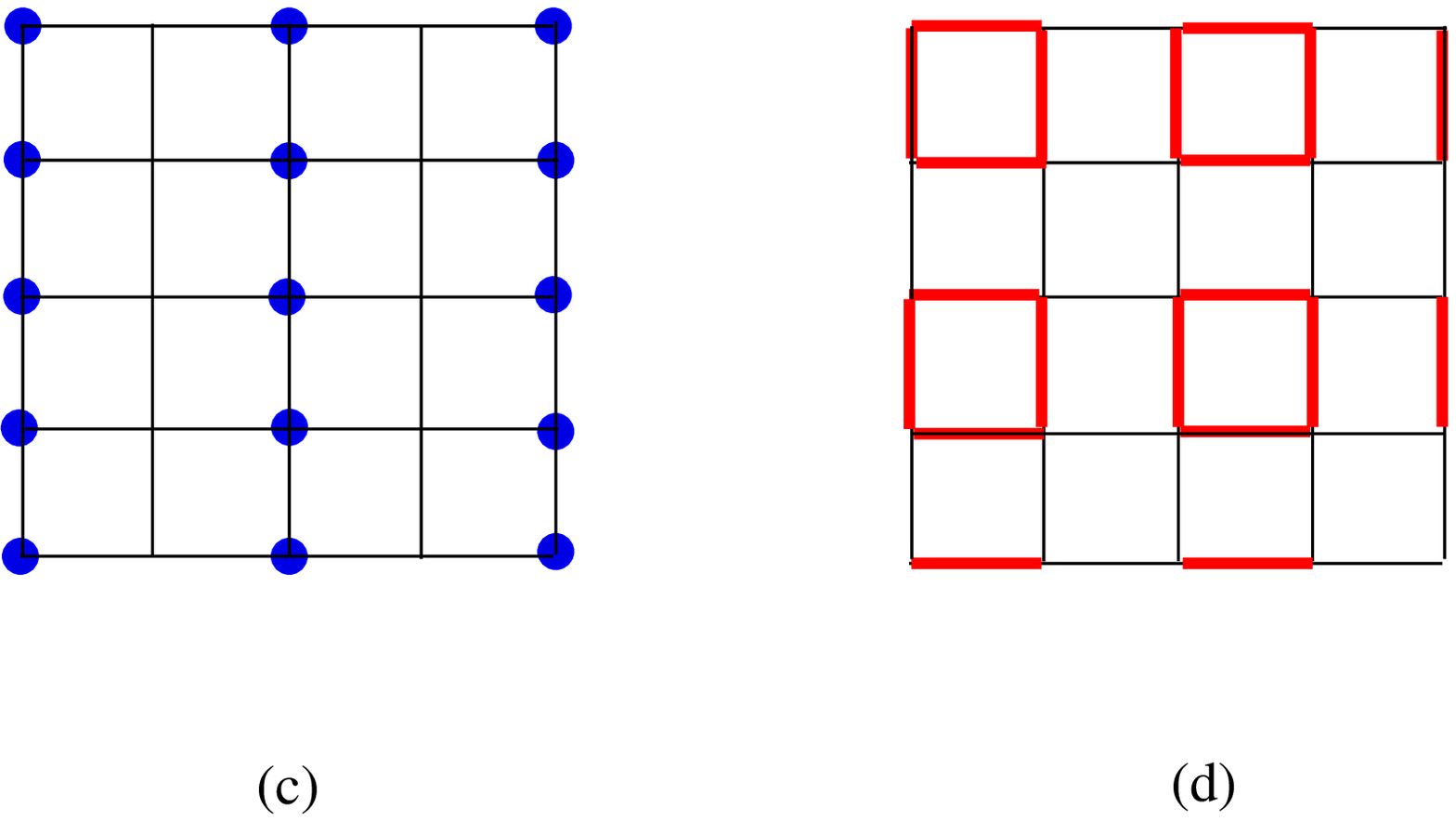}
\caption{ The charge density wave (CDW) phase in a square lattice at
$ n_0=1/2 $ with ordering wavevector $ \vec{Q}_{n}=(\pi,\pi) $. (b)
valence bond solid (VBS) phases with ordering wavevector $
\vec{Q}_{K}=(\pi,0) $ where the kinetic energy $ \langle K_{ij}
\rangle =  \langle b^{\dagger}_{i} b_{j} + h.c. \rangle $ takes a
non-zero constant $ K $ in the two sites connected with a dimer, but
$ 0 $ in the two sites without a dimer. (c) Stripe CDW order at  $
\vec{Q}_{n}=(\pi, 0) $ and (d) Plaquette VBS order at $
\vec{Q}_{n}=(\pi, 0), ( 0, \pi ) $.
\cite{boson,gan,square,frusqmc,pq1,mob,bipart,frus,honey}. }
\label{fig2}
\end{figure}




   The interaction between the two laser beams in Fig.1 with the two level bosonic atoms is:
\begin{eqnarray}
  H_{int} & = & \int d^{2} \vec{r} \Psi^{\dagger}(\vec{r}) [
  \frac{\vec{p}^{2}}{ 2 m_{a}} + V_{OL}( \vec{r} ) + \frac{ \hbar \omega_{a} }{2} \sigma_{z}    \nonumber  \\
   & + & \frac{\Omega}{2} \sum_{l} ( e^{-i \omega_{l} t } \sigma^{+} u_{l}(\vec{r}) + h.c. ) ] \Psi(\vec{r})
\label{bragg}
\end{eqnarray}
   where $ \Psi(\vec{r})=( \psi_{e}, \psi_{g} ) $ is the two component boson annihilation operator,
   the incident and scattered lights in Fig.1a and the two incident lights in Fig.1b have frequencies $ \omega_{l} $ and mode functions
   $ u_{l}( \vec{r} ) = e^{i \vec{k}_{l} \cdot \vec{r} + i \phi_{l} } $ .
   The Rabi frequencies $ \Omega $ are much
   weaker than the laser beams ( not shown in Fig.1 ) which form the optical lattices.
   When it is far off the resonance, the laser light-atom detunings $ \Delta_{l}= \omega_{l}
  -\omega_{a} $ where $ \omega_a $ is the two level energy  difference are much larger than the Rabi frequency $ \Omega $ and
  the  energy transfer $ \omega= \omega_1-\omega_2 $ ( See Fig.1a and 1b ), so $ \Delta_{1} \sim \Delta_{2} = \Delta $.
  After adiabatically eliminating the upper level $ e $ of the two level atoms,
  expanding the ground state atom field operator $ \psi_{g}(\vec{r}) = \sum _{i} b_{i} w( \vec{r} -\vec{r}_{i} )  $ in Eqn.\ref{bragg} where
  $ w( \vec{r} -\vec{r}_{i} ) $ is the localized Wannier functions of the lowest Bloch band corresponding to $ V_{OL}( \vec{r} ) $ and $  b_i $
  is the annihilation operator of an atom at the site $ i $ in the Eqn.\ref{boson}, then we get the effective interaction between the off-resonant
  laser beams and the ground level $ g $:
\begin{equation}
  H_{int} =  \hbar \frac{ \Omega^{2} }{ \Delta } e^{-i \omega t } [ \sum^{N}_{i}  J_{i,i} n_{i}
                   + \sum^{N}_{<ij>}  J_{i,j}   b^{\dagger}_{i}b_{j} ]
\label{laserint}
\end{equation}
  where the interacting matrix element is $  J_{i,j}= \int d \vec{r} w( \vec{r}-\vec{r}_{i} )
  u^{*}_{1}(\vec{r} ) u_{2}(\vec{r} ) w( \vec{r}-\vec{r}_{j} ) = J_{j,i} $.
   The first term in Eqn.\ref{laserint} is the on-site term $ \hat{D}=\sum^{N}_{i}  J_{i,i} n_{i} $ ( See Fig.1a ).
   The second term is the off-site term ( See Fig.1b ).
   Because the Wannier wavefunction $ w( \vec{r} ) $ can be taken as real in the
   lowest Bloch band, the off-site term can be written as $
   \hat{K} =  \sum^{N}_{<ij>}  J_{i,j} b^{\dagger}_{i}b_{j}
    = \sum^{N}_{<ij>}  J_{i,j} ( b^{\dagger}_{i}b_{j} + h.c.) $ which
   is nothing but the off-site coupling to the nearest neighbor kinetic energy of the bosons
   $ K_{ij}=   b^{\dagger}_{i}b_{j} + h.c. $.


    It is easy to show that:
\begin{equation}
  \hat{D} ( \vec{q} ) = f_{0}( \vec{q} ) \sum^{N}_{i=1} e^{-i \vec{q} \cdot \vec{r}_{i} } n_{i}= N f_{0}( \vec{q} ) n( \vec{q}  )
\end{equation}
  where $ \vec{q}= \vec{k}_{1} -\vec{k}_{2} $, $ f_{0}( \vec{q} ) = \int d \vec{r} e^{-i \vec{q} \cdot \vec{r} } w^{2}( \vec{r} )$  and
  $ n( \vec{q} ) = \frac{1}{N} \sum^{N}_{i=1} e^{-i \vec{q} \cdot \vec{r}_{i} } n_{i} =
  \sum_{\vec{k}} b^{\dagger}_{\vec{k}} b_{\vec{k} + \vec{q} } $ is the Fourier transform of the  density operator at
   the  momentum $ \vec{q} $. Note that $ n( \vec{q} )= n( \vec{q}+ \vec{K} )
   $. The wavevector is confined to $ L^{-1} < q < a^{-1} $ where the trap size
  $ L \sim 100 \mu m $ and the lattice constant $ a \sim 0.5 \mu m $ in Fig.1.
   In fact, more information is
   encoded in the off-site kinetic coupling in Eqn.\ref{laserint}.
    In a square lattice, the bonds are either oriented along the
    $ \hat{x} $ axis  $ \vec{r}_{j} - \vec{r}_{i}= \hat{x} $ or
    along the $ \hat{y} $ axis  $ \vec{r}_{j} - \vec{r}_{i}= \hat{y} $, we have:
\begin{equation}
  \hat{K}_{\Box}  =  N [  f_{x}( \vec{q} )   K_{x}( \vec{q} ) +
                f_{y}( \vec{q} ) K_{y}( \vec{q}  )]
\label{kxy}
\end{equation}
   where
   $  K_{\alpha}( \vec{q} )= \frac{1}{N} \sum^{N}_{i=1} e^{-i\vec{q} \cdot \vec{r}_{i} } K_{i,i+\alpha} =
   e^{ i q_{\alpha}/2} \sum_{\vec{k}} \cos k_{\alpha} b^{\dagger}_{\vec{k}} b_{\vec{k} + \vec{q} }   $
   are the Fourier transform  of the kinetic energy operator $ K_{ij} = b^{\dagger}_{i} b_{j} + h.c.
   $ along $ \alpha=x, y $ bonds at the  momentum $ \vec{q}  $
   and the "form" factors
   $ f_{\alpha}( \vec{q} )=f( \vec{q},  \vec{r}_{i} - \vec{r}_{j}=\alpha )=
   \int d \vec{r} e^{-i \vec{q} \cdot \vec{r} } w( \vec{r} ) w( \vec{r} + \vec{r}_{i} -\vec{r}_{j}
   ) $. Note that $ K_{\alpha}( \vec{q} )= K_{\alpha}( \vec{q}+ \vec{K} ) $.
   Following the harmonic approximation used in \cite{boson},
   we can estimate that   $  f_{0}( \pi,0  ) \sim e^{ -\frac{1}{4} ( V_0/E_r)^{-1/2} },
   f_{x}( \pi,0  ) \sim i e^{ -\frac{1}{4} ( V_0/E_r )^{-1/2} -\frac{\pi^{2}}{4} ( V_{0}/E_{r} )^{1/2} } $,
   so  $  | f_{x} ( \pi,0  )/f_{0}( \pi,0  ) | \sim e^{-\frac{\pi^{2}}{4} \sqrt{ V_{0}/E_{r}} }  $
   where $ V_{0} $ and $ E_{r} = \hbar^{2} k^{2}/2m $ are the strength of the optical lattice potential and
   the recoil energy respectively \cite{boson}. The $ f_{0}( \pi,0  ) $ is close to 1 when $ V_{0}/E_{r} > 4 $.
   It is instructive to relate this ratio to
   that of the hopping $ t $ over the onsite interaction $ U $ in the Eqn.\ref{boson}: $ | f_{x} ( \pi,0 )/f_{0}( \pi,0 ) |
   \sim \frac{t}{U} \frac{a_s}{a} $ where $ a_s $ is the zero field scattering length and $ a= \lambda/2= \pi/k $ is the lattice constant,
   using the typical values $ t/U \sim 10^{-1}, a_s/a \sim 10^{-2} $, one can estimate $ | f_{\alpha}/f_{0} | \sim 10^{-3} $.
   Note that the harmonic approximation works well only in a very deep optical lattice $ V_0 \gg E_r $, so the above value {\sl underestimates }
   the ratio, so we expect $ | f_{\alpha}/f_{0} | \ge 10^{-3} $.

   The differential scattering cross section of the light from the cold atom systems in the Fig.1 can be
   calculated by using the standard linear response theory:
\begin{eqnarray}
  \frac{ d \sigma }{ d \Omega dE } ={\cal S}( \vec{q}, \omega )   & \sim  & ( \frac{ \Omega^{2} }{ \Delta } )^{2}
  N^{2} [ |f_{0}( \vec{q} ) |^{2} S_{n}( \vec{q}, \omega )
  \nonumber  \\
  & + &  \sum_{\alpha=\hat{x},\hat{y} } |f_{\alpha}( \vec{q} ) |^{2} S_{K_{\alpha} }( \vec{q}, \omega ) ]
\label{linear}
\end{eqnarray}
     where  $ \vec{q}=\vec{k}_{1}-\vec{k}_{0}, \omega= \omega_{1}-\omega_{2} $,
     the $ S_{n}( \vec{q}, \omega ) = \langle n(-\vec{q}, - \omega ) n(\vec{q}, \omega ) \rangle $
     is the dynamic density-density response function whose Lehmann representation was listed in \cite{braggbog}.
     The $ S_{K_{\alpha}}( \vec{q}, \omega ) = \langle K_{\alpha}(-\vec{q}, - \omega ) K_{\alpha}(\vec{q}, \omega ) \rangle $
     is the  bond-bond response function whose Lehmann representation can be got from that of the $ S_{n}( \vec{q}, \omega ) $
     simply by replacing the density operator $ n( \vec{q} ) $ by the bond operator $  K_{\alpha}( \vec{q} ) $.
     The integrated scattering cross section over the
     final energy  $ \frac{ d \sigma }{ d \Omega } = \int d
     E  \frac{ d \sigma }{ d \Omega d E } $ is
     proportional to the {\sl equal-time} response function $   \frac{ d \sigma }{ d \Omega } = {\cal S}( \vec{q} ) \sim  ( \frac{ \Omega^{2} }{ \Delta } )^{2} N^{2} [  |f_{0}( \vec{q} ) |^{2} S_{n}( \vec{q}
     )  +  \sum_{\alpha=\hat{x},\hat{y} } |f_{\alpha}( \vec{q} ) |^{2} S_{K_{\alpha}}( \vec{q}) ] $.

    We first look at the superfluid to Mott transition at integer filling factor $ n $.
    When $ \vec{q} $ is equal to the shortest reciprocal lattice vector $ \vec{K} = ( 2 \pi, 0 ) $,
    in the Mott state, $  \frac{ d \sigma^{M} }{ d
    \Omega } \sim  | f^{M}_{0}( 2 \pi, 0) |^{2} N^{2} n^{2} $,  in the superfluid state, $  \frac{ d \sigma^{SF} }{ d
    \Omega } \sim  | f^{SF}_{0}( 2 \pi, 0) |^{2} N^{2} n^{2}  + 2 | f^{SF}_{x}( 2 \pi, 0) |^{2} N^{2} B^{2} $
    where $ B $ is the average kinetic energy on a bond in the
    superfluid side. Because $ | f^{SF}_{0}( 2 \pi, 0) |^{2} \sim | f^{M}_{0}( 2 \pi, 0) |^{2} \sim 1 $ and
    $ B $ is  appreciable in the superfluid side,
    we expect a dramatic increase of the scattering cross section
\begin{equation}
      \frac{ d \sigma^{SF} }{ d
    \Omega }- \frac{ d \sigma^{M} }{ d
    \Omega } = 2 | f^{SF}_{x}( 2 \pi, 0) |^{2} N^{2} B^{2}
\end{equation}
     across the Mott to the SF transition due to the prefactor $ N^{2}$. This prediction could be tested immediately.
    Surprisingly, there is no such optical Bragg scattering experiment in the superfluid yet.

   In the  CDW with $ \vec{Q}_{n} = (\pi,\pi ) $ in Fig.2a,
   due to the lack of VBS order on both sides, the second term in Eqn.\ref{linear} can be neglected, so that
\begin{equation}
   \frac{ d \sigma }{ d \Omega dE } |_{CDW} \sim ( \frac{ \Omega^{2} }{ \Delta } )^{2} N^{2} |f_{0}( \vec{q} ) |^{2} S_{N}( \vec{q}, \omega )
\label{linearcdw}
\end{equation}
   which should show a peak at $ \vec{q}= \vec{Q}_{n} $ ( Fig.3a ) whose amplitude
   scales as the {\sl square} of the number of atoms inside the trap
   $ \sim  |f_{0}(\pi,\pi ) |^{2} N^{2} m^{2}  $ where $ m= n_A-n_B $ is the CDW order parameter \cite{bipart}.
   When $ \vec{q} =\vec{K} $, then  $ {\cal S}_{CDW}( \vec{K} ) \sim  |f_{0}(2 \pi, 0 ) |^{2} N^{2} n^{2} $
   where $ f_{0}(2 \pi, 0 ) \sim f^{2}_{0}(\pi,\pi ) $ ( Fig.3a ).
   So the ratio of the two peaks in Fig.3a is $ \sim m^{2}/n^{2} $ if one
   neglects the very small difference of the two form factors.
   Slightly away from $ 1/2 $ filling, the CDW in Fig.2a may turn into the
   CDW supersolid ( CDW-SS ) phase through a second order phase transition \cite{bipart}. Then we have $
   \langle n(\vec{q}) \rangle = m  \delta_{\vec{q},\vec{Q}_{n}} +  n \delta_{\vec{q},0}  $ where $ n= n_A+n_B = 1/2 + \delta n $.
   The superfluid density $ \rho_s \sim \delta n= n -1/2 $.
   The scattering cross section inside the CDW-SS: $  {\cal S}_{CDW-SS}( \vec{Q}_n ) \sim |f_{0}( \pi, \pi ) |^{2} N^{2}
   m^{2} $ stays more or less the same as that inside the CDW, but $  {\cal S}_{CDW-SS}( \vec{K} ) \sim  |f_{0}(2 \pi, 0 ) |^{2} N^{2}
   n^{2} + 2 |f_{x}(2 \pi, 0 ) |^{2} N^{2} (\delta n)^{2} B^{2} $ will increase. The $ B $ is
   the average bond strength due to very small superfluid component  $ \rho_s \sim \delta n= n -1/2 $
   flowing through the whole lattice. So the right peak in
   Fig.3a will increase due to the increase of the total density and the superfluid component inside the CDW-SS phase.




  Now we discuss the VBS state  with $ \vec{Q}_{K}=(\pi,0) $ in Fig.2b. Due to the {\sl uniform } distribution of the density in the VBS,
  when  $ \vec{q}=\vec{K} $, the
  second term in Eqn.\ref{linear} can be neglected, so there is a diffraction peak ( Fig.3b ) whose amplitude scales as the
  {\sl square} of the number of atoms inside the trap
  $ \sim |f_{0}(2\pi,0) |^{2} N^{2} n^{2} $ where $ f_{0}(2 \pi, 0 ) \sim  f^{4}_{0}(\pi, 0 ) $ and $ n=1/2 $ is the uniform
  density in the VBS state. However, when one tunes $ \vec{q} $ near
  $ \vec{Q}_{K} $, the first term in Eqn.\ref{linear} can be neglected, then
\begin{equation}
   \frac{ d \sigma }{ d \Omega dE } |_{VBS} \sim ( \frac{ \Omega^{2} }{ \Delta } )^{2} N^{2}
   \sum_{\alpha=\hat{x},\hat{y} } |f_{\alpha}( \vec{q} ) |^{2} S_{K_{\alpha} }( \vec{q}, \omega )
\label{linearvbs}
\end{equation}
  which should show a peak at $ \vec{q} = \vec{Q}_{K} $ signifying the VBS ordering at $ \vec{Q}_{K} $
  whose amplitude scales also as the {\sl square} of the number of atoms inside the trap
  $ \sim |f_{x}(\pi,0) |^{2} N^{2} K^{2} $ where $ K=K_{x}-K_{y} $ is the
  VBS order parameter \cite{bipart}.
  So the ratio of the VBS peak at $ \vec{q}= \vec{Q}_{K} $ over the uniform density
  peak at $ \vec{q}= \vec{K} $ is $ \sim K^{2}/n^{2} |f_{x}(\pi,0)/f_{0}(2\pi,0) |^{2} \geq 10^{-5} $.
  However, the smallness of $ |f_{x} |^{2} $ is compensated by the large number of atoms $ N \sim  10^{6} $,
  $  |f_{x} |^{2} N^{2} =  ( |f_{x} |^{2} N ) \times N  \sim N \sim 10^{6} $.
  Therefore, the Bragg scattering cross section from the VBS order is $ \geq 10^{-5} $ smaller than that
  at $ \vec{q} =\vec{K} $ at the same incident energy $ I_{in} $ ( Fig.3b ), but still $ \sim 10^{6} $ above the background,
  so very much visible in the current optical Bragg scattering experiments.
  Slightly away from $ 1/2 $ filling, the VBS may turn into VB Supersolid (VB-SS) through a second order transition \cite{bipart}.
  We have $ \langle K_{x}(\vec{q}) \rangle =  B \delta_{\vec{q},0} + K \delta_{\vec{q},\vec{Q}_{K}}  $
  and $ \langle n(\vec{q}) \rangle = ( \delta n + 1/2 ) \delta_{\vec{q},0}  $.
  The superfluid density $ \rho_s \sim \delta n= n -1/2 $.
   The scattering cross section inside VB-SS: $ {\cal S}_{VB-SS}( \vec{Q}_K ) \sim |f_{x}( \pi, 0 ) |^{2} N^{2}
   K^{2} $ stays more or less the same as that inside the VBS, but $ {\cal S}_{VB-SS}( \vec{K} ) \sim  |f_{0}(2 \pi, 0 ) |^{2} N^{2}
   n^{2} +   |f_{x}(2 \pi, 0 ) |^{2} N^{2} (\delta n )^{2}  B^{2}_{x} + |f_{y}(2 \pi, 0 ) |^{2} N^{2} (\delta n )^{2} B^{2}_{y} $
   where $ n = 1/2 + \delta n $ and the $ B_x, B_y $
   are  the average bond strengths along $ x $ and $ y $ due to very small superfluid component  $ \rho_s \sim \delta n= n -1/2 $
   flowing through the whole lattice. So the right peak in
   Fig.3b will increase due to the increase of the total density and the superfluid component inside the VB-SS phase.
  Very similarly, one can discuss the VBS order at $   \vec{q}= \vec{Q}_{K}= ( 0, \pi ) $. For the plaquette VBS order in Fig.2d, then one should be able to
  see the $ S_{K}( \vec{q} ) $ peaks at both $ ( \pi, 0 ) $ and $ (0,\pi) $. So the dimer VBS and the plaquette VBS can also be distinguished by
  the optical Bragg scattering.



\begin{figure}
\includegraphics[width=7cm]{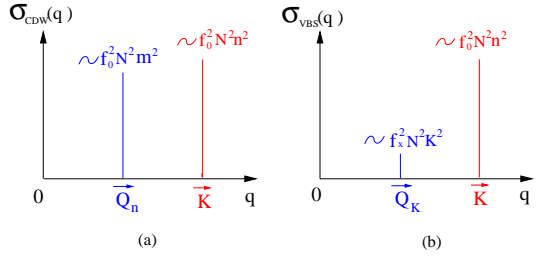}
\caption{ The  optical scattering cross section in (a) CDW, the
ratio of the peak at $ \vec{Q}_{n} $ over that at $ \vec{K} $ is $
\sim m^{2}/n^{2}\sim 1 $. (b) VBS state, the ratio of the peak at $
\vec{Q}_{K} $ over that at $ \vec{K} $ is $ \sim |f_{x}/f_{0}|^{2}
K^{2}/n^{2} \geq 10^{-5} $, but still should be visible in the
current optical Bragg scattering experiments. }
\label{fig3}
\end{figure}

       In this paper, we only focused on the optical Bragg scattering detections of the various ground
       states in a square lattice. The detections of the excitation spectra, the generalization to frustrated lattices,
       the effects of finite temperature and a harmonic trap  will be discussed in a future publication.

We thank G.G. Batrouni, Jason Ho, R.Hulet, S. V. Isakov, Juan Pino
and Han Pu for helpful discussions. J. Ye also thanks Jason Ho, A.
V. Balatsky and Han Pu for their hospitalities during his visit at
Ohio State, LANL and Rice university. J. Ye's research is supported
by NSF-DMR-0966413, at KITP is supported in part by the NSF under
grant No. PHY-0551164, at KITP-C is supported by the Project of
Knowledge Innovation Program (PKIP) of Chinese Academy of Sciences.
W.M. Liu's research was supported by NSFC-10874235. W.P. Zhang's
research was supported by the NSFC-10588402 and -10474055, the 973
Program under Grant No.2006CB921104.


\end{document}